\documentclass[english,aps,pre,amsmath,amssymb,showpacs,notitlepage,onecolumn]{revtex4-1}
\usepackage{amsmath}
\usepackage{mathtools}
\usepackage[svgnames]{xcolor}
\usepackage[colorlinks=true, linkcolor=Maroon, urlcolor=Blue]{hyperref} 
\usepackage{color}
\usepackage{float}
\usepackage{subfigure}
\usepackage{graphicx}
\hypersetup{
  colorlinks,
  citecolor=MidnightBlue,
  linkcolor=DarkRed,
  urlcolor=Blue}

\begin{document}
\title{Implementing Parrondo's paradox with two coin quantum walks}
\author{Jishnu Rajendran}
\author{Colin Benjamin}
\email{colin.nano@gmail.com}
\affiliation{School of Physical Sciences, National Institute of Science Education \& Research, HBNI, Jatni-752050,\ India }
\begin{abstract}
Parrondo's paradox is ubiquitous in games, ratchets and random walks.The apparent paradox, devised by J.~M.~R.~Parrondo, that two losing games $A$ and $B$ can produce an winning outcome has been adapted in many physical and biological systems to explain their working. However, proposals on demonstrating Parrondo's paradox using quantum walks failed {for large number of steps}. In this work, we show that instead of a single coin if we consider a two coin initial state which may or may not be entangled, we can observe a genuine Parrondo's paradox with quantum walks. Further we focus on reasons for this and pin down the asymmetry in initial two-coin state or asymmetry in shift operator, either of which are necessary for observing a genuine Parrondo's paradox. We extend our work to a  3-coin initial state too with similar results. The implications of our work for observing quantum ratchet like behavior using quantum walks is also discussed.   
\end{abstract}
\maketitle
\section{Introduction}
Parrondo's paradox consists of a sequence of games, individually each of which are losing games but provide a winning outcome when played in a deterministic or random order. It has been shown that Parrondo's games have important applications in many physical and biological systems\cite{control,gamethe}. Quantum version of Parrondo's games were introduced in Refs.[\onlinecite{flitney}-\onlinecite{Meyer}]. Quantum version of the classical random walk on other hand was introduced in 1993 in Ref.[\onlinecite{1993}] {
and is developed and studied extensively throughout the years\cite{Andraca}}. In Refs.\cite{flitney,Meyer,blumer} Parrondo's games are explored using 1-D discrete time quantum walk(DTQW). When a game is played, the net expectation of position of the walker defines a win or a loss. It has been already shown that quantum walk version of Parrondo's paradox does not exist in the asymptotic limits \cite{flitney,minli}. The need for studying Parrondo's games via quantum walks is necessitated by the search for applications in building better algorithms\cite{ken} and to explain physical process like quantum ratchets\cite{chandru2}.

\section{Motivation}
Our motivation in this work is to implement a genuine Parrondo's paradox via quantum walks. We show that while previous attempts at implementing Parrondo's paradox with quantum walks failed in the asymptotic limits\cite{flitney,minli} our method using two coin initial states gives a genuine Parrondo's paradox even in the asymptotic limits.
\\
Parrondo's game as originally introduced in Refs.\cite{Harmer,Parrondo} is a gambling game. A player plays against a bank with a choice of two games $A$ and $B$, whose outcomes are determined by the toss of biased coins. Each of these games is losing when played in isolation but when played alternately or in some other deterministic or random sequence (such as $ABB\ldots, ABAB\ldots$, etc.) can become a winning game. Owing to this counter-intuitive nature, Parrondo's games are also referred to as Parrondo's paradox. The apparent paradox that two losing games $A$ and $B$ can produce a winning outcome when played in an alternating sequence was originally devised by Juan M. R. Parrondo as a pedagogical illustration of the Brownian ratchet\cite{Parrondo}. Parrondo's games have important applications in many physical and biological systems, \emph{e.g.,} in control theory the random/deterministic combination of two unstable systems can produce a overall stable system\cite{control}.

\begin{figure}
\centering
\includegraphics[scale =0.5]{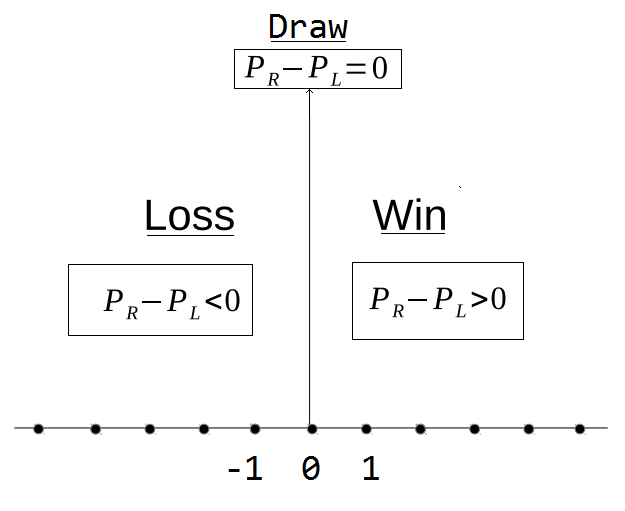}
\caption{Pictorial illustration of the conditions for win or loss for QWs on a line.}
\label{fig:win-loss}
\end{figure}

\begin{figure}[h]
  \centering \subfigure[]{ \includegraphics[width=0.43\textwidth]{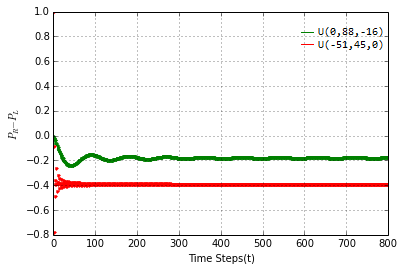}}
  \centering \subfigure[]{ \includegraphics[width=0.43\textwidth]{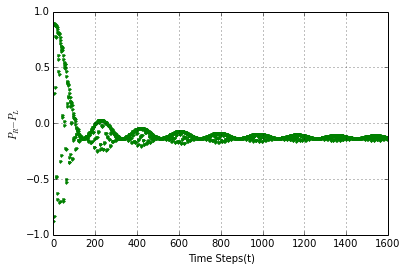}}
\caption{
 a) $P_{R}-P_{L}$ of the walker  after $t$ steps, with initial state $\vert\Psi_{0}\rangle=\frac{1}{\sqrt{2}}\vert 0\rangle\otimes(\vert 0\rangle-i\vert 1\rangle)$, and coin operator $A=U^{S}(-51,45,0)$ (red line) or $B=U^{S}(0,88,-16)$ (green line).  b) $P_{R}-P_{L}$ of the walker with games played in sequence $ABBBABBB\ldots$ (i.e., $q=4$), $A=U^{S}(-51,45,0)$, $B=U^{S}(0,88,-16)$ (1600 steps), herein initially you win($steps < 100$) but {at large steps} you lose.
}\label{game}
\end{figure}

The 1-D discrete time quantum walk(DTQW) implementation of Parrondo's paradox is as follows:
Consider two games A and B  played alternately in time. Game A and B are represented by different quantum operators $U(\alpha_{A},\beta_{A},\gamma_{A})$ and $U(\alpha_{B},\beta_{B},\gamma_{B})$\cite{chandru3,coin_control},

\begin{equation}
U(\alpha,\beta,\gamma)=\left(\begin{array}{cc}
e^{i\alpha}\cos\beta & -e^{-i\gamma}\sin\beta\\
e^{i\gamma}\sin\beta & e^{-i\alpha}\cos\beta
\end{array}\right).\label{eq:SU(2)}
\end{equation}
The initial state of the quantum walker is  $\vert\Psi_{0}\rangle=\frac{1}{\sqrt{2}}\vert 0\rangle\otimes(\vert 0\rangle-i\vert 1\rangle),$ where first ket refers to the position space and second ket refers to the single coin space which is initially in  a superposition of heads and tails. The shift in the position space, say from $|n\rangle$ to $|n-1\rangle$ or $|n+1\rangle$, is defined by a unitary operator called shift operator($\mathcal{S}$) defined as,
\begin{equation}
\mathcal{S} =   \sum\limits_{n=-\infty}^{\infty}\vert n+1 \rangle \langle n \vert \otimes \vert 0 \rangle \langle 0 \vert +   \sum\limits_{n=-\infty}^{\infty}\vert n-1 \rangle \langle n \vert \otimes \vert 1 \rangle \langle 1 \vert.
\label{Equ:S}
\end{equation}
 Games \emph{A} and \emph{B} are played alternately in different time steps, i.e., game \emph{A} is played on time steps
$t=nq$ and game \emph{B} is played on time steps $t\neq nq$, where $q$ is the period and $n$ is an integer. The evolution operator can be written as: 
\begin{equation}
U =\left\lbrace 
	\begin{array}{ll}
		\mathcal{S}\cdot U(\alpha_{A},\beta_{A},\gamma_{A})  & \mbox{if } t=nq,n\in Z \\
		\mathcal{S}\cdot U(\alpha_{B},\beta_{B},\gamma_{B}) & \mbox{if } t\neq nq,n\in Z
	\end{array}
\right. 
\end{equation}
and the final state after $N$ steps is given by $\vert\Psi_{N}\rangle=U^{N}\vert\Psi_{0}\rangle$. For $q=3$, it means we play games with the time sequence $ABBABB\ldots$. As denoted in Fig.~\ref{fig:win-loss}, after $N$ steps, if the probability $P_{R}$ of the walker to be found to the right of the origin, is greater than the probability $P_{L}$ to be found to the left of the origin, i.e., $P_{R}-P_{L}>0$, we consider the player to win. Similarly, if $P_{R}-P_{L}<0$, the player losses. If $P_{R}-P_{L}=0$, it means the player neither loses nor wins, it's a draw. By making use of the above scheme, Parrondo's games using 1-D DTQW are formulated. The game is constructed with two losing games \emph{A} and \emph{B} having two different biased coin operators $U_{A} (\alpha_{A},\beta_{A},\gamma_{A})$ and $U_{B}(\alpha_{B},\beta_{B},\gamma_{B})$, if we set $\alpha_A=-51,\beta_{A}=45,\gamma_{A}=0,\alpha_{B}=0,\beta_{B}=88,\gamma_B=0$,  $U_{A}^{S}=U^{S}(-51,45,0)$, $U_{B}^{S}=U^{S}(0,88,-16)$ as in Fig.~\ref{game}(a). We form a game with sequences \emph{$ABBB\ldots$}. This results in winning at the beginning but in the asymptotic limit the player will lose as in Fig.~\ref{game}(b), one can check for different sequences like \emph{$ABAB\ldots ABBABB\ldots$ etc.} and in all cases in the asymptotic limits we lose. Hence Parrondo's paradox does not exist in case of 1-D DTQW. This fact was noted in Refs.~\cite{flitney,minli} also. In particular, Ref.~\cite{flitney} shows with many different sequences like ABAB.., AABAAB.., etc, {at large steps} there is no Parrondo's paradox. Hence our motivation to find circumstances for the existence of a genuine Parrondo's paradox in quantum walks. {It is possible to show the convergence of quantum walk as obtained in Fig.~\ref{game}(a) analytically. The analytical form of convergence mentioned in Theroem 1 of Ref.\cite{konno} for a single coin quantum walk can be used for calculating the asymptotic limit.}
\begin{equation}
E=-(1-\sqrt{1-\vert e^{i\alpha }\cos\beta\vert^2}) \lambda
\end{equation}
where,
\begin{equation*}
\lambda = \frac{i e^{i\alpha}\cos\beta (-e^{i\gamma})\sin\beta + i e^{-i\alpha}\cos\beta (-e^{-i\gamma})\sin\beta}{2\vert e^{i\alpha\cos\beta}\vert^2}
\end{equation*}

{where $E$ is the convergence value of the quantum walk. The convergence for the single coin quantum walk when the coin operates $AAAA\ldots$ and $BBBB\ldots$( $A=U(-51,45,0)$,$B=U(0,88,-16)$) is calculated and found to be $-0.227621$ and $-0.480834$ respectively, which is very close to that of the numerical results found in Fig.~\ref{game}(a).}\\

 {Considering the classical limit of the quantum walk with a single coin, with classical operators Identity(I) and NOT(X) we obtain the classical results if $A=Identity$(I) and $B=NOT$(X) operators are used then the classical walk we have $P_R-P_L =0$ for $AAAA\ldots$ and  $BBBB\ldots$ as well as $ABAB\ldots$. This is in conformity with results of a classical random walk which has a gaussian distribution with mean as well as median equal to zero.}  


\section{Parrondo's paradox using two coin initial state}
As in the previous section, the elements of our two coin quantum walk are the walker, coins, evolution operators for both the coins, walker and a set of observables. The walker is a quantum system with its position denoted as 
$|\text{position}\rangle$ residing in a Hilbert space of infinite but countable dimension ${\cal H}_P$. The basis states $|i \rangle_P$ which span ${\cal H}_P$, and any superposition of the form $\sum_{i} \alpha_i|i\rangle_p$ which are subject to $\sum_i|\alpha_i|^2 = 1$, are valid states for the walker \cite{entangled}. The walker is usually initialized at the \lq origin', i.e., $|\text{position}\rangle_0 = |0\rangle_P$.
\begin{figure}
 \centering \subfigure[]{ \includegraphics[width=0.44\textwidth]{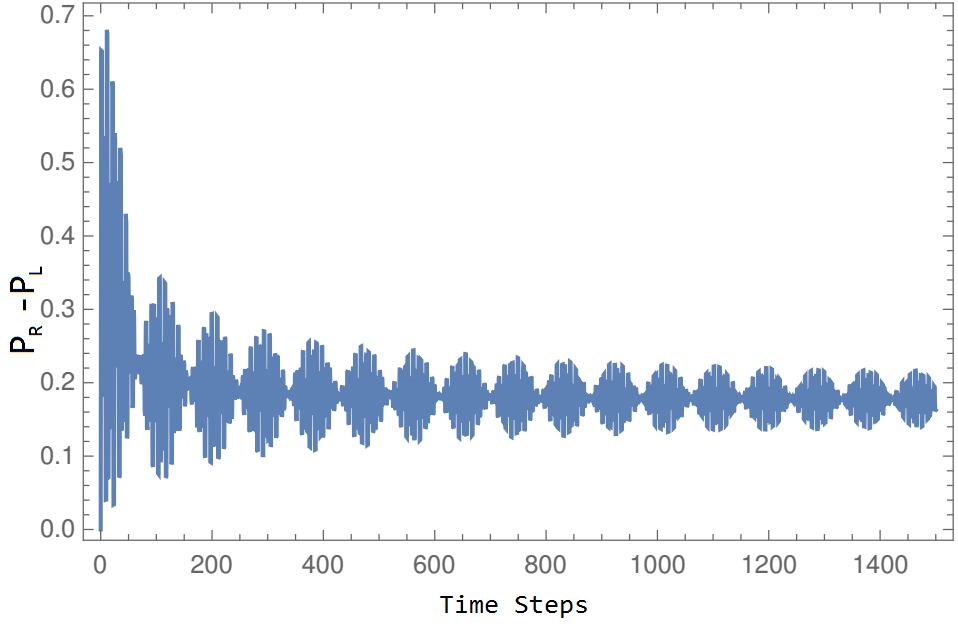}}
 \centering    \subfigure[]{ \includegraphics[width=.44\textwidth]{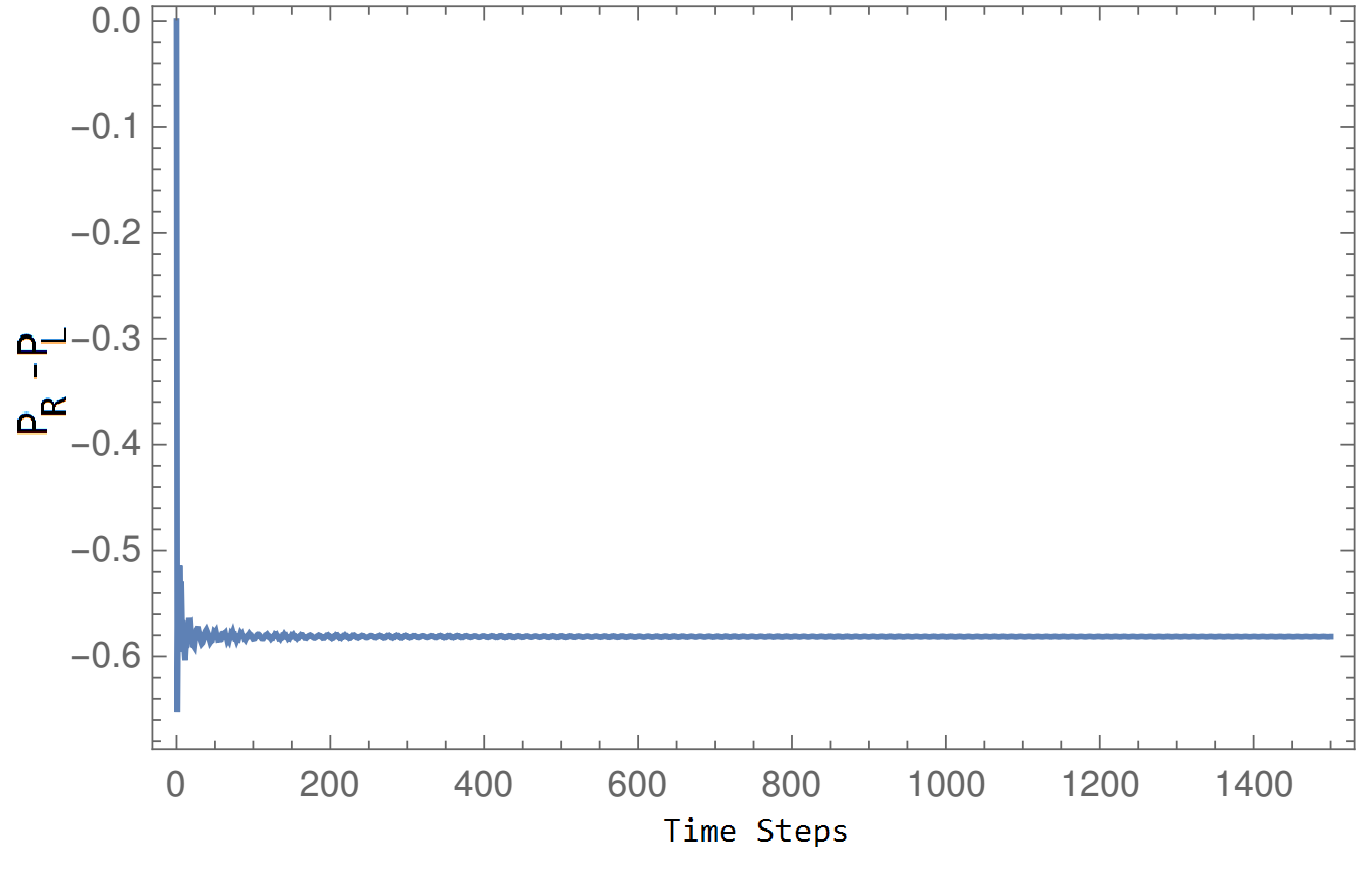}}
 \centering    \subfigure[]{ \includegraphics[width=.44\textwidth]{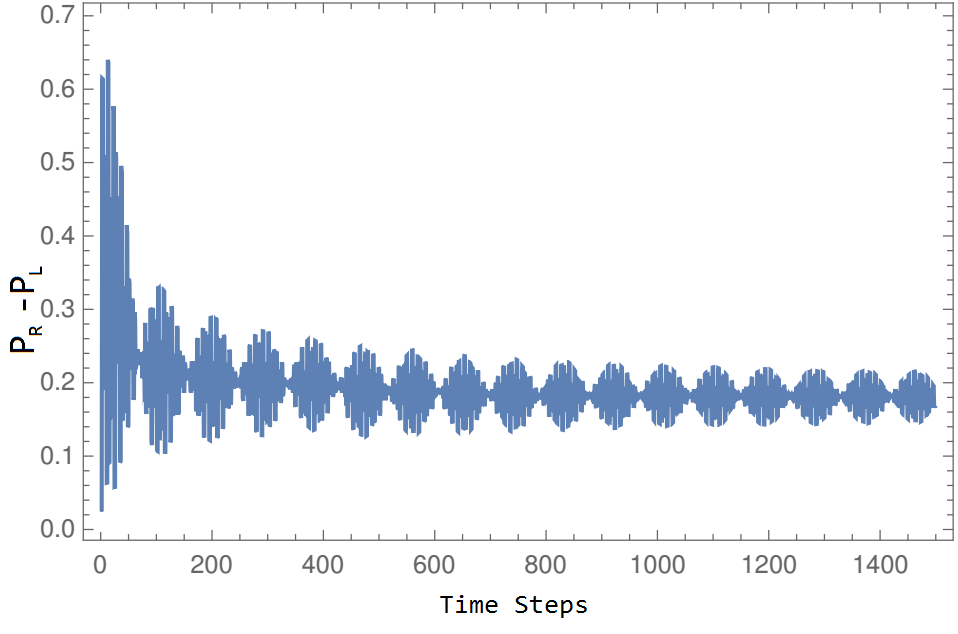}}
 \centering    \subfigure[]{ \includegraphics[width=.45\textwidth]{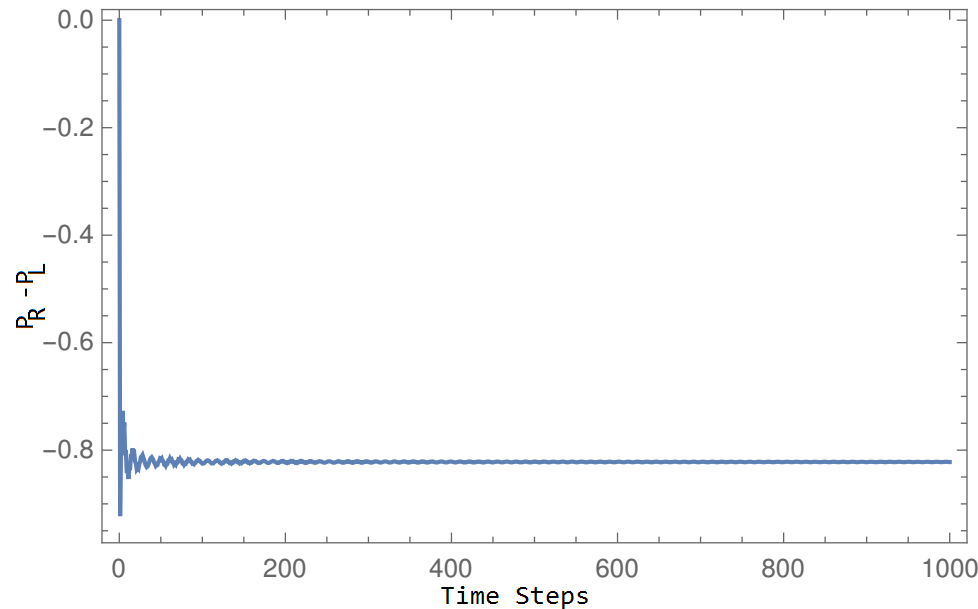}}
 \caption{a) Parrondo walk is evident even {at large number of steps} for partially entangled coin states ($\theta=\pi/4$) when $ABAB\ldots$ is played on first coin \& $BABA\ldots$ on second coin. b) However, when $AAAA\ldots$ is played on first \& $BBBB\ldots$ on second coin, one gets a losing outcome.  In c) we show similar to a partially entangled state a non-entangled state ($\theta=0$) also gives a Parrondo's paradox {for large number of steps} when $ABAB\ldots$ is played on first coin \& $BABA\ldots$ on second coin and finally in d) we show that $P_R-P_L$ is negative {at large steps} when $AAA\ldots$ and $BBB\ldots$ are played on the two coins.}\label{result}
\end{figure}
The two coin initial state is a quantum system in a 4-D Hilbert space ${\cal H}_{EC}$. We denote the two coin initial state as $|\text{coin}\rangle_0$, which may or may not be entangled-
\begin{equation}\label{e_state}
|\text{coin}\rangle_0 = \cos\left(\frac{\theta}{2}\right) |10\rangle + i \sin\left(\frac{\theta}{2}\right) |01\rangle.
\end{equation}
The initial state of the quantum walker resides in the Hilbert space ${\cal H}_T = {\cal H}_P \otimes {\cal H}_{EC}$ and has the form:
\begin{equation}
|\psi\rangle_0 = |\text{position}\rangle_0 \otimes
|\text{coin}\rangle_0
\end{equation}
which using Eq.~\ref{e_state}, gives $|\psi\rangle_0 = |0\rangle \otimes \left(\cos\left(\frac{\theta}{2}\right) |10\rangle + i \sin\left(\frac{\theta}{2}\right) |01\rangle \right)$.
Evolution operators used are unitary as before and since the coin is a bipartite system, the coin is defined as the tensor product of two single-qubit coin operators: $C_{EC}=U_{\alpha_k,\beta_k,\gamma_k}\otimes U_{\alpha_l,\beta_l ,\gamma_l}$, where $k$,$l$ can be any of the Game $A$ and $B$. The evolution operator is fully separable, thus any entanglement in the coins is due to the initial states used. The conditional shift operator $S_{EC}$  allows the walker to move either forward or backward, 
depending on the state of the coins. The operator 
\begin{equation}\label{shift_operator_two_spins}
S_{EC} =  \sum_i |i+1 \rangle_{pp} \langle i| \otimes |00\rangle_{cc} \langle 00|  
+ \sum_i |i\rangle_{pp} \langle i| \otimes |01\rangle_{cc} \langle 01| 
+ \sum_i |i\rangle_{pp} \langle i| \otimes |10\rangle_{cc} \langle 10|
+ \sum_i |i-1 \rangle_{pp} \langle i| \otimes |11\rangle_{cc} \langle 11|
\end{equation}
incorporates the stochastic behavior of the random walk with a two coin initial state. It is only when the coin is in the $|00\rangle$ or $|11\rangle$ state that the walker moves either forward or backward else the walker does not move.
The full evolution operator has the structure $U_T = S_{EC}.(I_p \otimes  C_{EC})$ and one can mathematically represent a two coin quantum walk after $N$ steps as $|\psi \rangle_N = (U_T)^N |\psi\rangle_0,$ where $|\psi\rangle_0$ denotes the initial state of the walker and the coins. As defined before, winning and losing in context of Parrondo's game, after $N$ time steps if the probability $P_R$ of the walker to be found to the right of the origin is greater than the probability to be found left of the origin, i.e., $P_R-P_L > 0$ we consider the player to win. However if, $P_R - P_L <  0$ then the player loses and if $P_R-P_L = 0$ it implies a draw. In order to obtain a genuine Parrondo's paradox the two games $A$ and $B$ are now played on the two  coin space as follows: $U_A \otimes U_B$ is operated on the two coins and in the next step $U_B \otimes U_A$ is played on the two coins. Thus, for the first coin we have the series $ABAB\ldots$ while on the second coin we have $BABA\ldots$. The coin operators can as before be defined as-
\begin{eqnarray*}
X&=&A \otimes B =C_{EC}=U(-51,45,0) \otimes U(0,88,-16)\nonumber\\
Y&=&B \otimes A =C'_{EC}=U(0,88,-16) \otimes U(-51,45,0)
\end{eqnarray*}

and are played alternately in time, i.e, in sequence $XYXY\ldots$ and the plot for $P_R -P_L$ as shown in Fig.~\ref{result}(a) is obtained. It is evident that the sequence $XYXY\ldots$ provides a winning outcome for two losing games {at large number of steps}. The fact that individually the sequence $AAA\ldots$ on first coin and $BBB\ldots$ on second coin give a losing outcome can be seen from $P_R-P_L$ plot in Fig.~\ref{result}(b).

\begin{figure}
\centering 
\includegraphics[width=.63\textwidth]{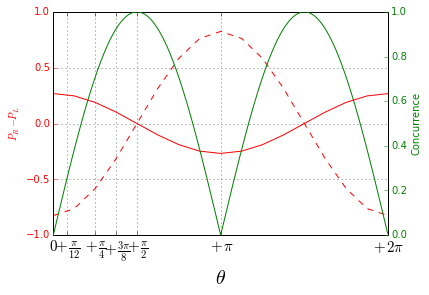}
\caption{Plot of Concurrence(Green), $P_{R}-P_{L}$ (red, solid) for ABAB.. on first coin and BABA...on second coin, and finally $P_{R}-P_{L}$ (red, dashed) for AAAA.. on first coin and BBBB...on second coin. Note that Parrondo's paradox is observed for $0<\theta<\pi/2$ and $3\pi/2<\theta<2\pi$ with the definition as in Fig.~\ref{fig:win-loss}. In the region $\pi/2 <\theta < 3\pi/2$ there is a role reversal and thus our definition for Parrondo's paradox as used in Fig.~\ref{fig:win-loss} is also reversed.}
\label{concur}
\end{figure}
\section{Discussion}
From Fig.~\ref{result} one can convincingly conclude that to obtain a genuine Parrondo's paradox via quantum walks one needs a non-entangled or a partially entangled two coin state. When a single coin was considered (as in Fig.~\ref{game}) the outcome of Parrondo's games did not give rise to the paradox {for quantum walk with large number of steps}. In order to obtain a Parrondo's paradox, what is needed is a two-coin state.  
\begin{figure}
\centering 
\includegraphics[width=.793\textwidth]{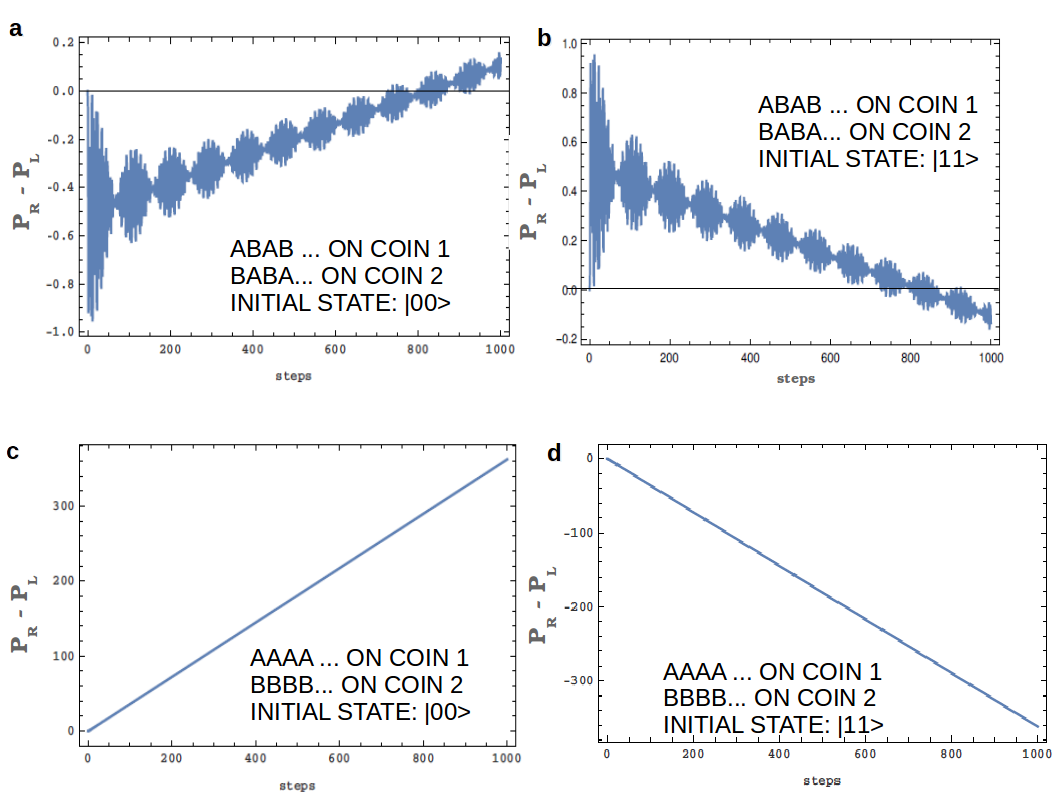}
\caption{Parrondo's paradox and the initial 2-coin state. a) Plot of  $P_{R}-P_{L}$ for ABAB.. on first coin and BABA...on second coin for state $|00\rangle$, b) Plot of  $P_{R}-P_{L}$ for ABAB.. on first coin and BABA...on second coin for state $|11\rangle$, c) Plot of  $P_{R}-P_{L}$ for AAAA.. on first coin and BBBB...on second coin for state $|00\rangle$ and finally d) Plot of  $P_{R}-P_{L}$ for AAAA.. on first coin and BBBB...on second coin for state $|11\rangle$. For both $|00\rangle$ as well as $|11\rangle$ state there is no Parrondo's paradox. For state $|00\rangle$ there is a role reversal and thus our definition for Parrondo's paradox as used in Fig.~\ref{fig:win-loss} is also reversed. }
\label{initial-fig}
\end{figure}
Finally, what are the plausible reasons for the success of the two coin initial state as compared to the single coin state? We can start by identifying the reasons which do not lead to Parrondo's paradox. First, entanglement has no or marginal role.  Maximally entangled coins lead to a draw as the probability distribution is perfectly symmetric as noted before in Ref.\cite{entangled}, on the other hand non-entangled or partially entangled coins lead to a Parrondo's paradox. Further, in Fig.~\ref{concur}, we plot the amount of entanglement present in a quantum system, i.e., the concurrence\cite{concurrence}. The concurrence is zero for a separable state and one for a maximally entangled state. Fig.\ref{concur} shows the concurrence for our arbitrary two coin state as a function of $\theta$. One sees that Parrondo's paradox is observed for $0<\theta<\pi/2$ and $3\pi/2<\theta<2\pi$ with the definition as in Fig.~\ref{fig:win-loss}. In the region $\pi/2 <\theta < 3\pi/2$ there is a role reversal and thus our definition for Parrondo's paradox as used in Fig.~\ref{fig:win-loss} is also reversed.

 Next, the initial state? Here the answer is more complicated. As we have seen, the initial state can be a product state for maximal violation of the paradox, this is evident from Fig.~\ref{result}. The initial state thus, does play a role, however this is not without qualification. The shift operator also plays a role which we discuss after this. 
\begin{figure}
\centering 
\includegraphics[width=.793\textwidth]{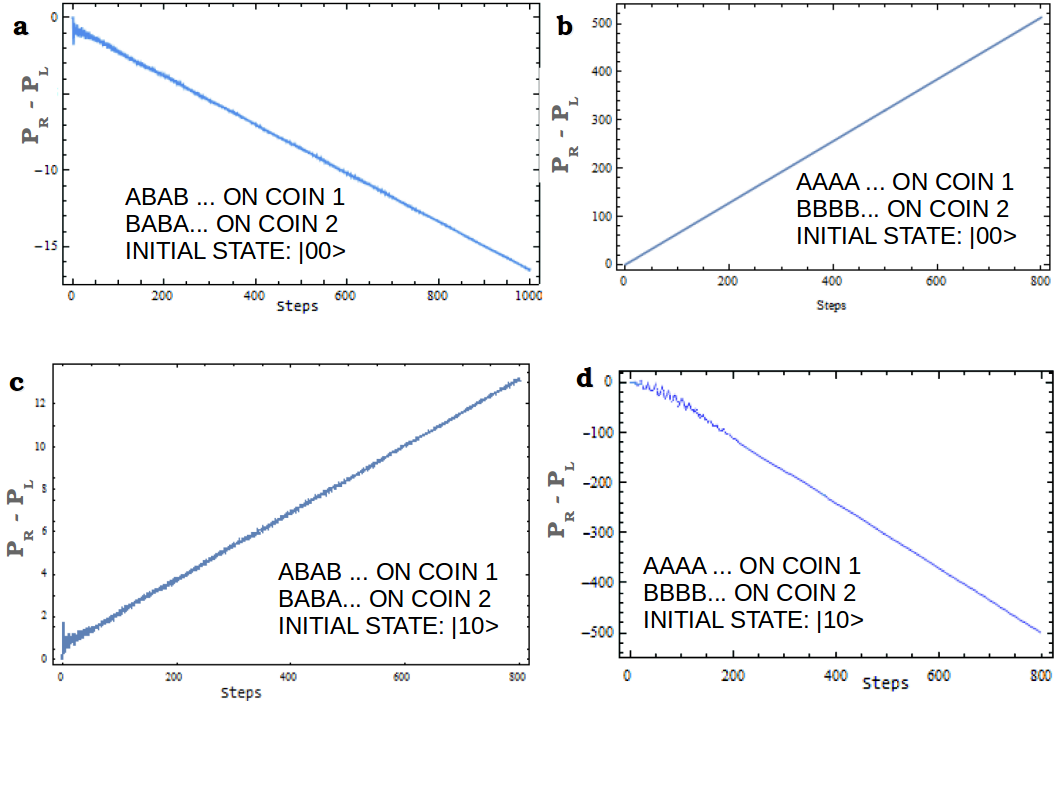}
\caption{ Parrondo's paradox with a shift operator with one wait state (Eq.~\ref{shift-one-wait}). a) Plot of  $P_{R}-P_{L}$ for ABAB.. on first coin and BABA...on second coin for state $|00\rangle$, b) Plot of  $P_{R}-P_{L}$ for {AAAA..} on first coin and {BBBB..} on second coin for state $|11>$, c) Plot of  $P_{R}-P_{L}$ for {ABAB..} on first coin and {BABA..} on second coin for state $|00\rangle$ and finally d) Plot of  $P_{R}-P_{L}$ for AAAA.. on first coin and BBBB.. on second coin for state $|11\rangle$. For both $|00\rangle$ as well as $|11\rangle$ state there is now Parrondo's paradox with shift operator as defined in Eq.~\ref{shift-one-wait}. For state $|00\rangle$ there is a role reversal and thus our definition for Parrondo's paradox as used in Fig.~\ref{fig:win-loss} is also reversed. }
\label{shift-fig}
\end{figure}
\begin{figure}
\centering 
\includegraphics[width=.793\textwidth]{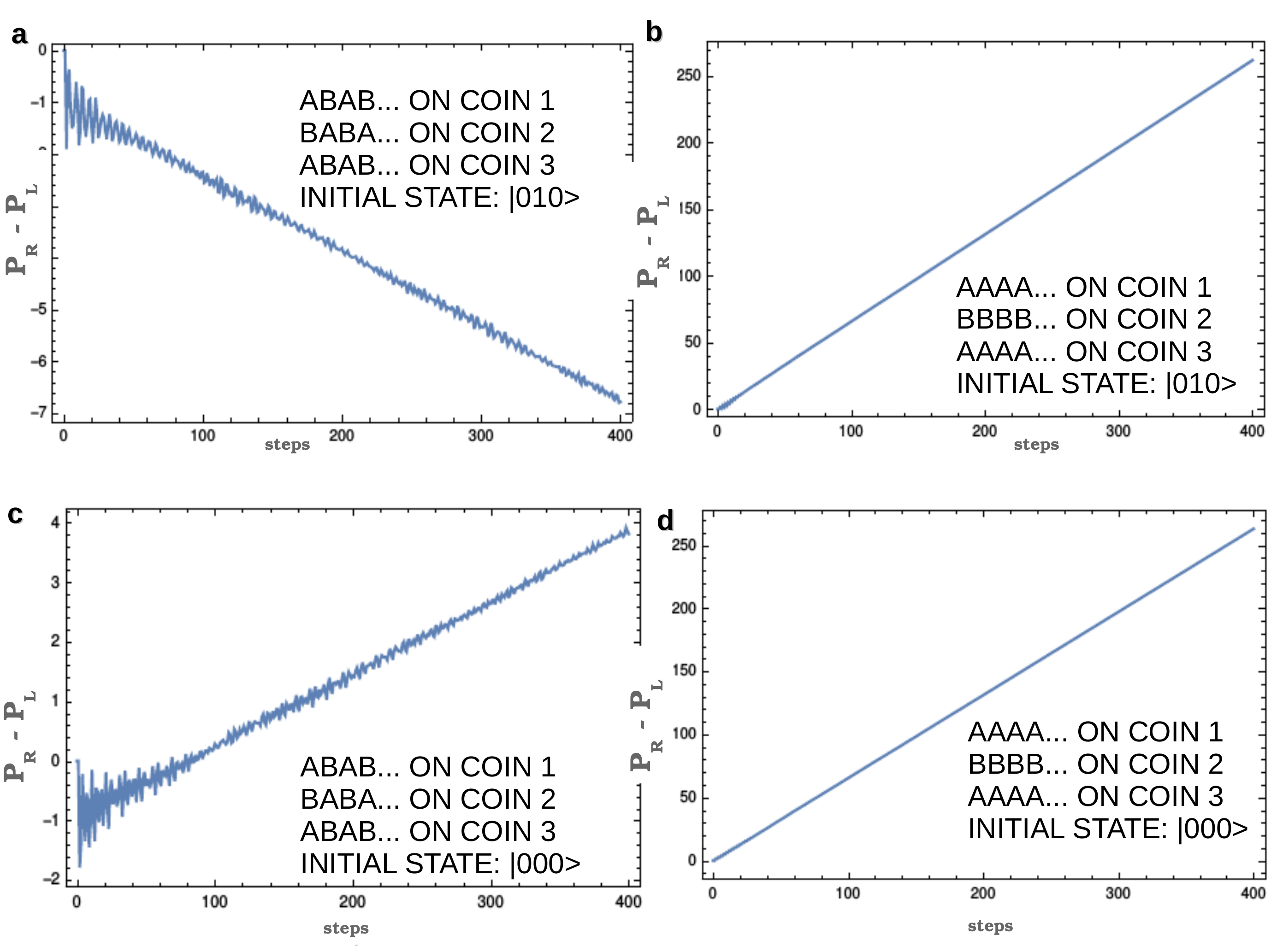}
\caption{ Parrondo's paradox with an initial 3-coin state. a) Plot of  $P_{R}-P_{L}$ for ABAB.. on first and third coins and BABA...on second coin for state $|010\rangle$, b) Plot of  $P_{R}-P_{L}$ for AAAA.. on first and third coins and BBBB...on second coin for state $|010\rangle$, c) Plot of  $P_{R}-P_{L}$ for ABAB.. on first coin and BABA...on second coin for state $|000\rangle$ and finally d) Plot of  $P_{R}-P_{L}$ for AAAA.. on first and third coins and BBBB...on second coin for state $|000\rangle$. For  $|000\rangle$ state there is no Parrondo's paradox. For state $|010\rangle$ we see a Parrondo's paradox however there is a role reversal and thus our definition for Parrondo's paradox as used in Fig.~\ref{fig:win-loss} is also reversed. }
\label{result3}
\end{figure}
Lets focus on the initial state. Supposing the shift operator is defined as  before in Eq.~\ref{shift_operator_two_spins}. We compare the quantum walks starting with initial states $|11\rangle$ and $|00\rangle$ in Fig.~\ref{initial-fig}. Both do not lead to the Parrondo's paradox. This may give the impression that only when one has initial 2-coin state $|01\rangle$ or $|10\rangle$ composed of orthogonal coin states do we see a Parrondo's paradox. However, it's not the complete picture. The shift operator plays a non-trivial role. If we change the shift operator, see Eq.~\ref{shift_operator_two_spins} from two wait states to just a single wait state as in Eq.~\ref{shift-one-wait} then a different picture emerges.
\begin{equation}
\label{shift-one-wait}
S_{EC} =  \sum_i |i+1 \rangle_{pp} \langle i| \otimes |00\rangle_{cc} \langle 00|  
+ \sum_i |i+1\rangle_{pp} \langle i| \otimes |01\rangle_{cc} \langle 01| 
+ \sum_i |i\rangle_{pp} \langle i| \otimes |10\rangle_{cc} \langle 10|
+ \sum_i |i-2 \rangle_{pp} \langle i| \otimes |11\rangle_{cc} \langle 11|
\end{equation}

In Fig.~\ref{shift-fig}, we plot $P_R - P_L$ for around $800$ time steps for both initial states- $|00\rangle$ as well as $|10\rangle$ with the new shift operator defined with a single wait state as in Eq.~\ref{shift-one-wait}. In this case for both $|00\rangle$ and $|10\rangle$ states we see Parrondo's paradox. To conclude the most plausible reason for observing the Parrondo's paradox is both due to some asymmetry which comes into play in a two coin state and is not possible to include in the single coin state.  The asymmetry may be in the initial quantum state or in the shift operator. 

Finally, what are the implications for more than two coin initial state? To test this we consider two different  3 coin initial states: $|010\rangle$ and $|000\rangle$. Similar to the 2 coin case discussed earlier, the coin is defined as the tensor product of three single-qubit coin operators: $C_{EC}=U_{\alpha_k,\beta_k,\gamma_k}\otimes U_{\alpha_l,\beta_l ,\gamma_l}\otimes U_{\alpha_m,\beta_m ,\gamma_m}$, where $k$,$l$ and $m$ can be any of the games $A$ and $B$. The conditional shift operator $S_{EC}$  allows the walker to move either forward or backward, depending on the state of the coins and is defined as-
\begin{eqnarray}
\label{shift_operator_three_spins}
S_{EC} &=&  \sum_i |i+2 \rangle_{pp} \langle i| \otimes |000\rangle_{cc} \langle 000|  
+ \sum_i |i+1\rangle_{pp} \langle i| \otimes |001\rangle_{cc} \langle 001| 
+ \sum_i |i\rangle_{pp} \langle i| \otimes |010\rangle_{cc} \langle 010|
+ \sum_i |i \rangle_{pp} \langle i| \otimes |011\rangle_{cc} \langle 011|\nonumber\\
&+&\sum_i |i \rangle_{pp} \langle i| \otimes |100\rangle_{cc} \langle 100|  
+ \sum_i |i\rangle_{pp} \langle i| \otimes |101\rangle_{cc} \langle 101| 
+ \sum_i |i-1\rangle_{pp} \langle i| \otimes |110\rangle_{cc} \langle 110|
+ \sum_i |i-2 \rangle_{pp} \langle i| \otimes |111\rangle_{cc} \langle 111|\nonumber\\
\end{eqnarray}
incorporates the stochastic behavior of the random walk with a three coin initial state. When the coin is in the $|000\rangle$ or $|111\rangle$ state that the walker moves two steps at once either forward or backward, while when the coin is in state $|001\rangle$ or $|110\rangle$ the walker moves one step either forward or backward and for the rest of the cases the walker remains fixed.
The full evolution operator similar for the two coin case has the structure $U_T = S_{EC}.(I_p \otimes  C_{EC})$ and the three coin quantum walk after $N$ steps is written as $|\psi \rangle_N = (U_T)^N |\psi\rangle_0,$ where $|\psi\rangle_0$ denotes the initial state of the walker and the coins. In order to obtain a genuine Parrondo's paradox the two games $A$ and $B$ are now played on the three coin space as follows: $U_A \otimes U_B\otimes U_A$ is operated on the three coins and in the next step $U_B \otimes U_A \otimes U_B$ is played on the three coins. Thus, for the first and third coins we have the series $ABAB\ldots$ while on  second coin we have $BABA\ldots$. The coin operators can as before be defined as-
\begin{eqnarray*}
X&=&A \otimes B \otimes A =C_{EC}=U(-51,45,0) \otimes U(0,88,-16) \otimes U(-51,45,0)\nonumber\\
Y&=&B \otimes A \otimes B =C'_{EC}=U(0,88,-16) \otimes U(-51,45,0) \otimes U(0,88,-16)
\end{eqnarray*}
and are played alternately in time, i.e, in sequence $XYXY\ldots$ and the plot for $P_R -P_L$ as shown in Fig.~\ref{result3}(a) is obtained. It is evident that the sequence $XYXY\ldots$ provides a winning outcome for two losing games even {at large number of steps}. The fact that individually the sequence $AAA\ldots$ on first and third coins while $BBB\ldots$ on second coin gives a losing outcome can be seen from $P_R-P_L$ plot in Fig.~\ref{result3}(b). In Fig.~\ref{result3}(c) and (d) we plot the median $P_R - P_L$ for the initial 3-coin state $|000\rangle$, we confirm the absence of any Parrondo's paradox for this initial state, confirming the trend seen for 2-coin initial states with symmetric shift operator.

To conclude, this section the initial state has a great bearing on having the Parrondo's paradox in a quantum walk or not. In both the two coin and three coin state when coins are orthogonal we see the paradox and for the case when they are not paradox disappears. Of course the aforesaid is subject to the qualification that the shift operator which controls the position of the coin state has an important bearing. Asymmetry in either the initial two coin state, e.g., $|10\rangle$ or $|01\rangle$ or an asymmetric shift operator (in case the initial state is $|00\rangle$ or $|11\rangle$) is necessary for obtaining a Parrondo's paradox with quantum walks.
\section{Conclusion}
Our goal in this work was to show evidence of a genuine Parrondo's paradox using quantum walks and we show this using a two coin state. We also considered entanglement between the two coins and showed that maximally entangled states do not show any paradox while non-entangled as well as partially entangled states do show the paradox. We also tried to understand the reasons behind this paradox. The most plausible reason behind observing the paradox with two or higher coin initial states is the introduction of  asymmetry either in the initial coin state or in the shift operator with one or two wait states in addition to left or right shifts. Our work can be considered as a demonstration of a  quantum ratchet too, implying particle transport against an applied bias in presence of noise or perturbations. In our case the noise parameter can be considered to reduce entanglement, thus looking at Fig.~\ref{concur}, from zero asymmetry in probability distribution, i.e., non-directed transport, when there is maximal entanglement to finite asymmetry in probability distribution, i.e., directed transport when there is no entanglement, is a clear marker of quantum ratchet like behavior of our system. The quantum ratchet analogies in Parrondo's paradox with quantum walks were also noticed in Ref.~\cite{Meyer}, however without any entanglement. New quantum walks are of great interest to the community as their investigation may lead to new quantum algorithms, which are of great interest to the quantum computation community at present.

 \section{Data Accesability} ``This article has no additional data''. The Mathematica code used in our paper has been adapted from the Mathematica examples on quantum walks available in the public domain in Ref.\cite{package}, see for additional examples http://homepage.cem.itesm.mx/lgomez/quantum/
\section{Competing Interests}
We have no competing interests.
\section{Author Contributions}
C.B. conceived the proposal, J.R. did the calculations on the advice of C.B., J.R. and C.B. wrote the paper, J.R. and C.B analyzed the results.

\section{Funding}
SCIENCE \& ENGINEERING RESEARCH BOARD, NEW DELHI, GOVT. OF INDIA  funded this research under GRANT NO. EMR/2015/001836.

\section{Research Ethics} This study did not require any prior ethical assesment.

\section{Animal Ethics} This study did not require any prior ethical assessment

\section{Permission to carry out fieldwork} No permissions were required prior to conducting this research.

\acknowledgements
C.B. thanks SCIENCE \& ENGINEERING RESEARCH BOARD, NEW DELHI, GOVT. OF INDIA for funding this research under GRANT NO. EMR/2015/001836.

\end{document}